\documentclass[12pt]{article}
\pdfoutput=1
\usepackage{epsf}
\usepackage{epsfig}
\usepackage{graphics}
\usepackage{graphicx}
\usepackage{amssymb}

\newcommand{\nc}{\newcommand}
\nc{\beq}{\begin{equation}}   \nc{\eeq}{\end{equation}}
\nc{\bea}{\begin{eqnarray}}   \nc{\eea}{\end{eqnarray}}
\nc{\baa}{\begin{array}}      \nc{\eaa}{\end{array}}
\nc{\bit}{\begin{itemize}}    \nc{\eit}{\end{itemize}}
\nc{\ben}{\begin{enumerate}}  \nc{\een}{\end{enumerate}}
\nc{\bce}{\begin{center}}     \nc{\ece}{\end{center}}
\nc{\non}{\nonumber}

\begin{document}
\rightline{November 2013}

\rightline{CUMQ/HEP-166}

\rightline{UQAM-PHE/01-2013}

\vspace{30pt}

\centerline{\Large \bf
Partial $\mu-\tau$ Textures and Leptogenesis}

\vspace{1cm}
\centerline{\large \bf Cherif  Hamzaoui$^{(a)}$\footnote{hamzaoui.cherif@uqam.ca},
Salah  Nasri$^{(b, c)}$\footnote{snasri@uaeu.ac.ae} and Manuel  Toharia$^{(d)}$\footnote{mtoharia@physics.concordia.ca} }

\begin{center}
\vspace{.4cm}
  \begin{it}

$^{(a)}$ Groupe de Physique Th\'eorique des Particules,\\
D\'epartement des Sciences de la Terre et de L'Atmosph\`ere,\\
Universit\'e du Qu\'ebec \`a Montr\'eal, Case Postale 8888, Succursale. Centre-Ville,\\
 Montr\'eal, Qu\'ebec, Canada, H3C 3P8.\\

\vspace{0.2cm}

$^{(b)}$  Department of Physics,United Arab Emirates University,\\
  P.O.Box 17551, Al-Ain, United Arab Emirates.\\
\vspace{0.2cm}

$^{(c)}$ Laboratoire de Physique Theorique,\\
Universit\'e d'Oran,  Es-S\'enia ,  DZ-31000, Oran,  Algeria.\\
 \vspace{0.2cm}

$^{(d)}$ Department of Physics, Concordia University\\
7141 Sherbrooke St. West, Montreal, Quebec, Canada.\\

\vspace{0.2cm}
\end{it}
\end{center}

\vspace{1.cm}
\centerline{\large \bf Abstract}

Motivated by the recent results from Daya Bay, Reno and Double Chooz
Collaborations, we study the consequences of small departures from
exact $\mu-\tau$ symmetry in the neutrino sector, to accomodate a
non-vanishing value of the element $V_{e3}$ from the leptonic
mixing matrix. Within the see-saw framework, we identify simple
patterns of Dirac mass matrices that lead to approximate
$\mu-\tau$ symmetric neutrino mass matrices, which are consistent with the
neutrino oscillation data and lead to non-vanishing mixing angle
$V_{e3}$ as well as precise predictions for the CP violating
phases. We also show that there is a transparent link between neutrino mixing angles and
see-saw parameters, which we further explore within the context of
leptogenesis as well as double beta decay phenomenology.\newpage

\section{Introduction}

Neutrinos are among the most elusive particles of the Standard Model
(SM) as they mainly interact through weak processes. Nevertheless, a clear
picture of the structure of the lepton sector has emerged thanks to the
many succesful neutrino and collider experiments over the past
decades.
The leptonic mixing angles,
contrary to the quark mixing angles are large. In fact, the very
recent results from T2K \cite{T2K}, Double Chooz \cite{DChooz},
RENO \cite{RENO} and Daya Bay \cite{DBay} Collaborations confirm that
even the smallest of the observed mixing angles, $\theta_{13}$, of the neutrino mixing
matrix is not that small.

We start this work with the observation that the data from
neutrino oscillations seem to show an approximate symmetry between the
second and third lepton families, also referred to as $\mu-\tau$
symmetry \cite{MUTAU, MUTAU2} (see also \cite{Altarelli}).
Exact $\mu-\tau$ symmetry when implemented at the level of the Majorana
neutrino mass matrix $S_{\nu}$, leads to the following relations
between its elements, namely $S_{12}=S_{13}$ and $S_{22}=S_{33}$. This
special texture of $S_{\nu}$ as well as different types of corrections
to it have been studied largely in the literature
\cite{literature}. Exact $\mu-\tau$ implemented in the charged lepton
basis is also known to lead, among other possibilities, to a vanishing
mixing angle $V_{e3}$ and a maximal atmospheric mixing angle
$|V_{\mu3}|=\frac{1}{\sqrt{2}}$.

We would like to put forward some
simple deviations from exact $\mu-\tau$ textures for $S_{\nu}$ in the context of the
simple see-saw mechanism \cite{seesaw}, and we call these $partial$ $\mu-\tau$ textures.
To do so we follow a bottom-up approach and construct textures for the Dirac neutrino mass
matrix $M_{D}$ in the limit in which we relax one of the two previous relations
coming from the exact $\mu-\tau$ symmetry.
Our main goal is to investigate if a small deviation from exact $\mu-\tau$ symmetry is
sufficient to generate the whole mixing structure in the lepton
sector, including ${\cal CP}$ violation, consistent with the existing
experimental data on neutrino oscillations.

We also require that the elements of the light neutrino
mass matrix $S_{\nu}$ and the Dirac neutrino mass matrix $M_{D}$ to be
independent.
As a consequence, we obtain a few allowed
simple textures for the Dirac neutrino mass matrix $M_{D}$ which in
turn leads to simple textures for the light neutrino mass
matrix.
Among the few possibilities allowed, we single out a simple texture
and study fully its phenomenological consequences.
In particular the chosen texture prefers an
inverted spectrum for the three active neutrinos and predicts the
value of the Dirac CP violating phase $\delta_D$. The impact of such
type of textures on leptogenesis and neutrinoless double beta decay will then be
considered as well as the associated relationship between low energy
and high energy CP violating parameters \cite{lowcphighcp}.

\section{Partial $\mu-\tau$ See-Saw}

We consider the most simple and popular mechanism for generating tiny
neutrino masses, namely the see-saw mechanism \cite{seesaw}.
In the case of Dirac neutrinos, the analysis is exactly the same as
quarks. However for the general case of Majorana neutrinos,
one obtains at low energies an effective mass matrix for the light left-handed Majorana
which is complex symmetric related to the Dirac mass matrix, $M_D$, as:
\begin{eqnarray}
S_{\nu}= - M_{D}M_{R}^{-1}M_{D}^{T}
\end{eqnarray}
We will work in the basis where the Majorana neutrino mass matrix $M_{R}$ is a diagonal
matrix. So we can parameterize its inverse as
$M_{R}^{-1}=\frac{1}{M_1} diag(1, R_{12},R_{13})$, with the Majorana
hierarchy ratios defined as $R_{12}=M_1/M_2\ $ and $\ R_{13}=M_1/M_3$.

To study the consequences of any symmetry implemented at the Lagrangian level in the
leptonic sector, it is instructive to construct a Dirac mass matrix $M_D$ which leads
naturally to a simple partial $\mu - \tau$ symmetric light neutrinos mass matrix $S_{\nu}$.
In general, $M_D$ is an arbitrary complex matrix:
\begin{eqnarray}
M_{D} =\left(\begin{array}{ccc}
 a &  b &  c\\
 d&  e&  f\\
 g &  h &  k\end{array}\right)
\end{eqnarray}
This gives us an $S_{\nu}$ of the form 
\begin{equation}
S_{\nu} =-\frac{1}{M_1} \left(\begin{array}{ccc}
a^2+R_{12}\ b^2+R_{13}\ c^2\ \ &  ad+R_{12}\ be+R_{13}\ cf\ \  &  ag+R_{12}\ bh+R_{13}\ ck\\
ad+R_{12}\ be+R_{13}\ cf\ \  &  d^2+R_{12}\ e^2+R_{13}\ f^2\ \ &  dg+R_{12}\ eh+R_{13}\ fk\\
ag+R_{12}\ bh+R_{13}\ ck\ \ &  dg+R_{12}\ eh+R_{13}\ fk\  \  &  g^2+R_{12}\ h^2+R_{13}\ k^2 \end{array}\right) .
\end{equation}
%
An exact $\mu-\tau$ texture happens when  $S_{22}=S_{33}$ and
$S_{12}=S_{13}$. This texture is known to have the $A_4$ and $D_4$
symmetry groups to be their possible underlying family symmetries
\cite{Ma2004,AltarelliA4}.
We therefore evaluate the differences between the elements of $S_\nu$,
$(S_{12}-S_{13})$, $(S_{22}-S_{33})$ as well as $(S_{22}-S_{23})$:
\begin{eqnarray}
\!\!S_{12}\!-\!S_{13}&=&\!\frac{1}{M_1}\! \Big[a(g-d)+R_{12} b(h-e)+R_{13} c(k-f)\Big]\ \label{1213}\\
\!\!S_{22}\!-\!S_{33}&=&\!\frac{1}{M_1}\! \Big[ (g^2-d^2)+R_{12} (h^2-e^2)+R_{13} (k^2-f^2)\Big]\ \ \ \ \label{2233}\\
\!\!S_{23}\!-\!S_{22}&=&\!\frac{1}{M_1}\! \Big[ d(d-g)+R_{12} e(e-h)+R_{13} f(f-k)\Big]  \ \label{2223}
\end{eqnarray}
From these equations we note that if we want to reproduce the neutrino
mass matrix $S_{\nu}$ in the limit of exact $\mu - \tau$
without forcing relations between the elements of the Dirac mass
matrix $M_D$ and those of the heavy Majorana neutrino mass $M_{R}$,
then we must have the second row of $M_D$ to be equal to its third
row, i.e
\begin{eqnarray}
g=d,\ \ \  h=e \ \ {\rm and} \ \  k=f. \label{mtmagic}
\end{eqnarray}
However this strong limit forces the determinant of $M_D$ to vanish
which in turn forces the determinant of $S_{\nu}$ to vanish also. This
means that at least one of the eigenvalues of $S_{\nu}$ must vanish.
This can also be understood from Eq.~(\ref{2223}) which
shows that the relations from Eq.~(\ref{mtmagic}) will produce
additional constraints on the symmetric neutrino mass matrix, quite
stronger than $\mu-\tau$ symmetry, namely $S_{12}=S_{13}$ and
$S_{22}=S_{33}=S_{23}$. The possibility of vanishing eigenvalues is
allowed by the data and has been studied by many authors
\cite{0212341,Ibarra,0602084}. Since this limit constrains strongly
our parameter space, we prefer to avoid it and remain as general as possible.

We will therefore consider small deviations from exact $\mu-\tau$ in
this see-saw context. In particular we would like to put forward
minimal textures for the Dirac mass matrix $M_D$ which maintain at
least one of the two $\mu-\tau$ constraints on $S_\nu$, i.e. either
$S_{12}=S_{13}$ is kept, with $S_{33} \neq S_{22}$, or $S_{22}=S_{33}$
is maintained with now $S_{13} \neq S_{12}$.
We call this type of setup {\it ``partial $\mu-\tau$''} as it maintains at least
one of the original $\mu-\tau$ constraints on the elements of the
neutrino mass matrix.
In the following, we will only consider the {\it
  ``partial $\mu-\tau$''} case  $S_{22}=S_{33}$ {\it and} $S_{13} \neq
S_{12}$, for a specific texture. A complete study of all possible
cases with many more examples will be presented elsewhere.

\section{Partial $\mu-\tau$ with $S_{22}=S_{33}$ and $S_{11}+S_{12} = S_{22}+S_{23}$}

By inspection of equations (\ref{1213}) and (\ref{2233}) we note that
to produce the desired  deviation from $\mu-\tau$, 
we have three natural textures 
which we dub texture I, texture II, and texture III respectively.
Each texture is associated with one of the eigenvalues of
$M_{R}^{-1}$, such that the breaking of exact $\mu-\tau$ symmetry is
proportional to $1$ for texture I, to $R_{12}$ for texture II, and to
$R_{13}$ for texture III:

\begin{eqnarray}
M_{D} = \left ( \matrix{a & b & c \cr d & e & f \cr -d & e & f }\right)
, \phantom{pp} M_{D}=\left ( \matrix{
 a & b & c \cr d & e & f \cr d
& -e & f }\right)
,\phantom{pp} M_{D}=\left ( \matrix{
a & b & c \cr d & e & f \cr d
& e & -f }\right).
\end{eqnarray}

Note the importance of the minus signs which break the degeneracy of
some of the entries, allowing the vanishing of $(S_{22}-S_{33})$, but
not that of $(S_{13}-S_{12})$.
Of course we are interested in small deviations from $\mu-\tau$
symmetry and this approach allows us to control these with the
Majorana mass hierarchy parameters $R_{12}$ or $R_{13}$. In the light of the recent results from
T2K \cite{T2K}, Double Chooz \cite{DChooz}, RENO \cite{RENO} and Daya
Bay \cite{DBay} Collaborations pointing out to a large $\theta_{13}$,
Texture I being the largest, by definition, and therefore becomes the natural starting
point of our study. So we will concentrate our attention to it in what follows.
If furthermore we implement the tri-bimaximal \cite{Harrison} condition, namely, $S_{11}+S_{12}=S_{22}+S_{23}$,
an interesting  texture emerges for the Dirac mass matrix $M_{D}$ which in turn gives us a very
simple form for $S_{\nu}$. Now by avoiding relations between the elements of the Dirac mass
matrix $M_D$ and those of the heavy Majorana neutrino mass $M_{R}$, we obtain two
interesting patterns for $M_D$ which satisfy $Det(M_D)\neq 0$.
Taking into account the above features, for instance, we obtain for the texture I
the following allowed two patterns:

\begin{eqnarray}
M_{D_1}^{I} =\left(\begin{array}{ccc}
a & b & c \\
-a & -\frac{b}{2} & c \\
a & -\frac{b}{2} & c \end{array}\right)
\end{eqnarray}

\begin{eqnarray}
M_{D_2}^{I} =\left(\begin{array}{ccc}
a & b & c \\
-a & b & -\frac{c}{2} \\
a & b & -\frac{c}{2} \end{array}\right)
\end{eqnarray}


\section{Example Case Study: Texture I}

We now concentrate on the phenomenology of the first special texture
emerging from Texture I. In particular, we start with
the following texture,

\begin{eqnarray}
M_{D_1}^{I} =\left(\begin{array}{ccc}
a & b & c \\
-a & -\frac{b}{2} & c \\
a & -\frac{b}{2} & c \end{array}\right)
\end{eqnarray}

Now we put forward a minimal texture for $M_{D}$ with
the additional requirements of non vanishing elements
$(M_D^{\dagger}M_D)_{12}$ (or $(M_D^{\dagger}M_D)_{13}$) and
$(M_D^{\dagger}M_D)_{11}$, necessary for successful leptogenesis as
well as non-vanishing determinant of $M_D$.
The goal is to keep the parameter content as minimal as possible while keeping the main features motivated
by the partial $\mu-\tau$ ansatz in order to fully describe the
neutrino masses, neutrino mixing and ${\cal CP}$ violation, as well as
the additional possibility of leptogenesis. Taking into account all of
this, we further simplify the previous texture by setting $c=b=m_D$ so
that  in the basis where $M_{R}$ is diagonal, we have the following
texture (and redefining $z=\frac{a}{b}$)
\begin{eqnarray}\label{MinText}
M_{D}^{I} = m_D\left(\begin{array}{ccc}
z & 1 & 1 \\
-z & -\frac{1}{2} & 1 \\
z & -\frac{1}{2} & 1 \end{array}\right)
\end{eqnarray}
where $m_D$ sets the Dirac mass scale and its phase is a global unphysical
phase. With this parametrization, the resulting light neutrino mass matrix $S_{\nu}$ is given by
\begin{eqnarray}
S_{\nu} = -\frac{2}{3} \tilde{m}_\nu \left(\begin{array}{ccc}
\varepsilon+\frac{(3+\eta_M)}{2} & -\varepsilon+\frac{\eta_M}{2} & \varepsilon+\frac{\eta_M}{2} \\
-\varepsilon+\frac{\eta_M}{2} & \varepsilon+\frac{(3+2\eta_M)}{4} & -\varepsilon+\frac{(3+2\eta_M)}{4} \\
\varepsilon+\frac{\eta_M}{2} & -\varepsilon+\frac{(3+2\eta_M)}{4} &
\varepsilon+\frac{(3+2\eta_M)}{4} \end{array}\right)
\label{Snu}
\end{eqnarray}
where we have introduced the parameters $\varepsilon$ and $\eta_M$ defined by
\bea
\varepsilon=\frac{M_2}{M_1}z^2 \hspace{1cm}
 {\rm and} \hspace{1cm}
M_{2}=\frac{M_{3}}{2}(1+\eta_M).
\eea
Both parameters will prove to be important in this ansatz,
and they both depend on the hierarchy between two heavy Majorana
masses. In particular the parameter $\eta_M$ denotes the deviation
from the special relationship $\displaystyle M_2=\frac{M_3}{2}$
between the two heaviest Majorana neutrino masses. Large deviations
from that special relationship will produce physical neutrino mass
splittings too large to be phenomenologically acceptable.

We have also defined the light neutrino mass scale $\tilde{m}_\nu$ as
\bea
\tilde{m}_\nu = \frac{3}{2}\frac{m_D^{2}}{M_2},
\eea
exemplifying the see-saw mechanism at work, since $m_D$ is an
electroweak scale mass parameter and $M_2$ is a heavy Majorana mass of
intermediate scale.

The matrix $S_{\nu}$ is diagonalized as:

\begin{eqnarray}
U_{\nu}^{\dagger}S_{\nu}U_{\nu}^{*}= D_{\nu}
\end{eqnarray}

where
\begin{eqnarray}
U_{\nu}=P_{L}V_{CKM}P_{R},
\end{eqnarray}

$P_{L}$ and $P_{R}$ are diagonal phase matrices and $V_{CKM}$ \cite{Cabibbo} is a
{\it CKM-like} mixing matrix with one phase and three angles which can
be parametrized as

\bea
V_{CKM-Like} =\left(\begin{array}{ccc}
 \ \  \times & |V_{e2}| &  |V_{e3}| e^{-i \delta_{D}} \\
 \ \  \times &\times &  |V_{\mu 3}|\ \ \ \ \ \ \\
 \ \  \times &\times  & \times \ \ \ \ \ \ \end{array}\right).
\eea

The phases in $P_{L}$ 
can be rotated away in the charged current basis, and the ones in $P_{R}=diag(1, e^{i\alpha}, e^{i\beta})$
describe Majorana ${\cal CP}$ violating phases. The $V_{PMNS}$ \cite{PMNS} mixing matrix is then given by:

\begin{eqnarray}
V_{PMNS}=V^{CKM-Like}P_{R}
\end{eqnarray}

We can now compute the determinant of $S_{\nu}$ in our
ansatz, and obtain the simple exact relation

\begin{eqnarray}
|{m}_1||{m}_2||{m}_3| = \frac{ 4}{3} |\tilde{m}_\nu|^3  (1+\eta_M) |\varepsilon|.
\end{eqnarray}
With it, we obtain approximate analytical expressions for the mixing
angles in the neutrino sector for small enough values of
$|\varepsilon|$ and $\eta_M$.
In particular, we find that
\begin{eqnarray}
V_{e3}&=&
-\frac{2\sqrt{2}}{3} |\varepsilon| e^{-i\theta_\varepsilon} \ \ +\ \ {\cal
  O} (|\varepsilon|^2)
\end{eqnarray}
and so, at this expansion order, we can trade the parameter $|\varepsilon|$ by the mixing
angle $|V_{e3}|$, and its phase $\theta_\varepsilon=Arg(z^2)$ is identified
as the dirac phase $\delta_D$, i.e. $\delta_D\simeq
\theta_\varepsilon$. We can now express the rest of the mixing
entries as expansions in powers of $|V_{e3}|$ and $\eta_M$.
We find
\begin{eqnarray}
|V_{\mu3}|^2 &\simeq&
\frac{1}{2}-\frac{1}{2}|V_{e3}|^2\
 +\ \ {\cal  O} \left(\eta_M|V_{e3}|,|V_{e3}|^3\right) \label{vmu3eq}\\
{\rm\hspace{-1cm}  and\hspace{2cm} \ \ \ }&&\nonumber\\
|V_{e2}|^2
&\simeq&\frac{1}{2}+\frac{1}{r}\left[\frac{|V_{e3}|}{\sqrt{2}}\cos{\delta_D}
  + \frac{5}{4}|V_{e3}|^2 -\frac{\eta_M}{3}\right] +\ \
 {\cal  O} \left(\eta_M|V_{e3}|,|V_{e3}|^3\right) \label{ve2}
\end{eqnarray}
where we  have introduced the neutrino mass hierarchy parameter $r$ given by
\bea
r=\frac{\Delta m^2_{21}}{\Delta m^2_{13}} = \frac{|m_2|^2-|m_1|^2}{|m_1|^2-|m_3|^2}
\eea
\begin{figure}[t]
 \center
\includegraphics[width=12.2cm,height=8.cm]{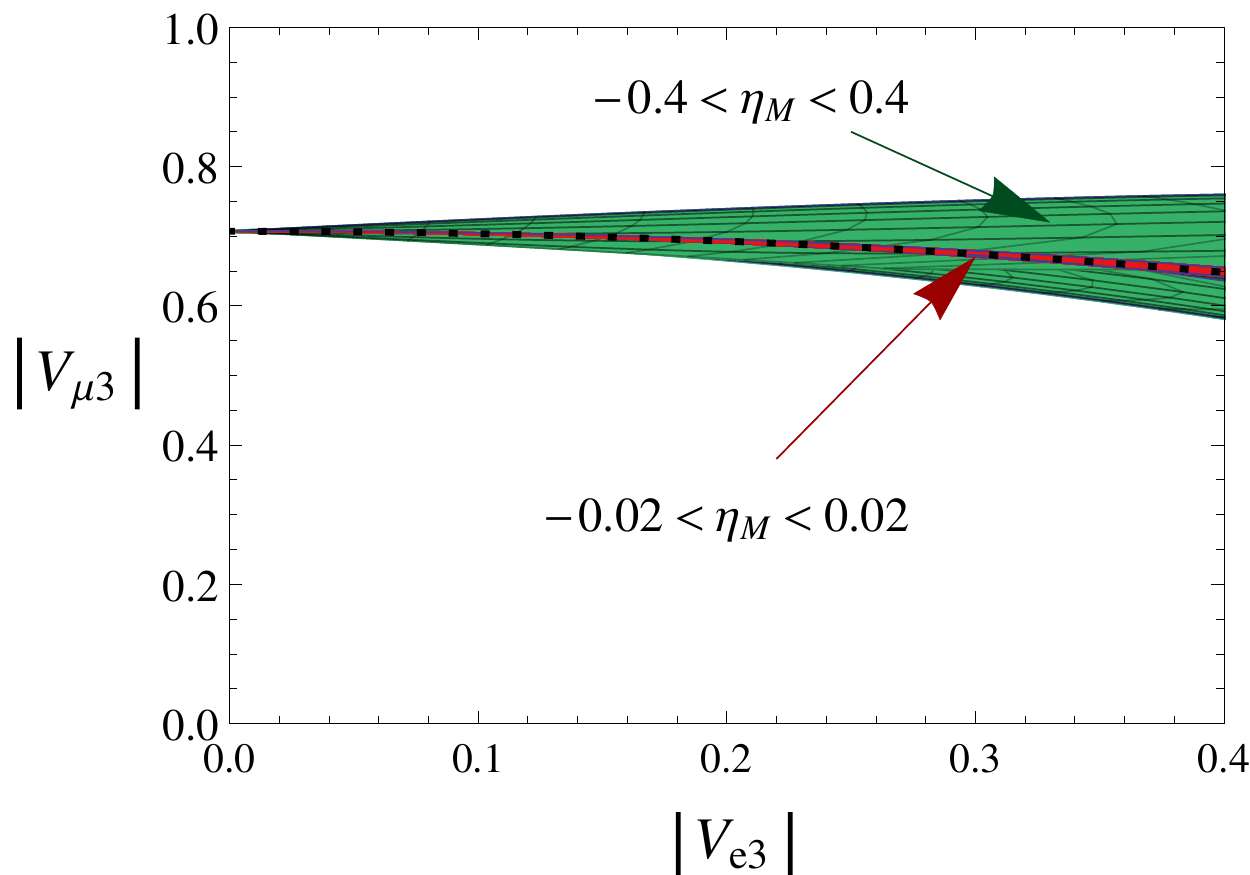}\hspace{1cm}
 \caption{Parametric plot of $|V_{\mu3}|$ with respect to $|V_{e3}|$
   varying $\eta_M$ from $-0.4<\eta_M<0.4$ in the large (blue) triangular shaded area, and
   $-0.02<\eta_M<0.02$ in the central thin (red) band (the region where
   acceptable $\Delta m^2_{21}$ can be obtained (see Figure
   \ref{contplot})). The dotted curve is the approximate expression
   obtained in Eq.~(\ref{vmu3eq}). The phase $\theta_{\varepsilon}\simeq\delta_D$ is here allowed the
   whole range from $0$ to $2\pi$, although its value fixes $|V_{e2}|$
   (see also Figure \ref{contplot}).}
\label{vmu3}
\vspace{.2cm}
 \end{figure}
As expected, the value of the atmospheric mixing angle is not far from
the exact $\mu-\tau$ symmetry value $|V^{0}_{\mu3}|^2=\frac{1}{2}$ with the
deviation being suppressed by the smallness of $|V_{e3}|^2$. Also
note that its value must lie in the first octant, i.e. the correction
is negative. We show in Figure 1 the numerical dependence of
$|V_{\mu3}|$ as a function of $|V_{e3}|$, allowing the Dirac phase $\delta_D$ to
take any value and limiting the possible values of $\eta_M$. The
simple analytical approximation of Eq.~(\ref{vmu3eq}) is also shown as a
dotted curve and it proves to be a very good approximation when the
values of $\eta_M$ are small, which as we will shortly see happens to be a
phenomenological requirement.

The physical neutrino masses predicted by the setup are such that
$|m_1|^2 \sim |m_2|^2 \sim |\tilde{m}_\nu|^2$ and
\bea
|m_3|^2 &\simeq&  2 |V_{e3}|^2\  |\tilde{m}_\nu|^2
\eea
so that the spectrum corresponds to an inverted mass hierarchy spectrum, and
the lightness of the lightest neutrino $\nu_3$ is explained by the
smallness of $|V_{e3}|$.
The solar neutrino mass $\Delta m^2_{21}=|m_2|^2-|m_1|^2$ is also small, but its expression is a complicated
admixture of terms of similar order in $\eta_M$, $|V_{e3}| \cos{\delta_D}$ and $|V_{e3}|^2$.

From Eq.~(\ref{ve2}) it might seem that for very small
$\eta_M$ and $|V_{e3}|$ the value of $|V_{e2}|^2$ approaches
$\frac{1}{2}$. This is not so, since the value of $r$ depends itself
on $\eta_M$ and $|V_{e3}|$. The limiting values for $|V_{e2}|^2$ are
\bea
\lim_{\eta_M\to 0}|V_{e2}|^2 &=& 1\ \ \ {or} \ \ \ 0\\
\lim_{|V_{e3}| \to 0}|V_{e2}|^2 &=& \frac{1}{3} \ \ \ \ \ (\eta_M >0) \label{goodlimit}\\
\lim_{|V_{e3}| \to 0}|V_{e2}|^2 &=& \frac{2}{3} \ \ \ \ \  (\eta_M
<0),
\eea
where the choice of $1$ or $0$ in the first limit depends on a flip of
masses $|m_1|$ and $|m_2|$ controlled by the value of $\delta_D$.
The experimentally preferred value of $|V_{e2}|$ is closest to the
limit of Eq.~(\ref{goodlimit}), meaning that the
model naturally produces it when $|V_{e3}|$ is sufficiently small and when $\eta_M$ is
positive. In that limit we have also
\bea
\lim_{|V_{e3}| \to 0} r = 2 |\eta_M|  ,
\eea
where $r=\frac{\Delta  m^2_{21}}{\Delta m^2_{13}}$, and in that
situation we see that the value of $\eta_M$ (which parameterizes the
deviation from the relationship $\displaystyle M_2=\frac{M_3}{2}\ $)
directly fixes the hierarchy measured between the neutrino mass differences, given by
$r_{exp}= \frac{{\Delta  {m^{2}_{21}}}_{exp}}{{\Delta m^2_{13}}_{exp}}
\simeq 0.03$
 (which would require that $|\eta_M| \sim 0.015$).\\

Of course, $|V_{e3}|$ does not seem to be so small according to the
recent reactor neutrino experiments results, with a
value sitting around $|V_{e3}|\sim 0.15$ according to global analysis
fits \cite{Valle,Fogli,GonzalezGarcia}.  For these larger values of $|V_{e3}|$, the parameters $\eta_M$,
$|V_{e3}|^2$ and/or $(|V_{e3}|\cos\delta_D)$ can be of the same
order and the (nice) tight prediction of $|V_{e2}|$ is lost, as it can now
take almost any value. In Figure 2 we show the
regions allowed by the experimental bounds on $|V_{e2}|$ (the blue
bands) and $r$ (the green ellipses), in terms of the Dirac phase
$\delta_D$ and the Majorana mass parameter $\eta_M$. The viable
regions (the intersections) are quite restricted and point towards small $\eta_M
\sim\pm 0.015$ and pretty well constrained values of $\delta_D$.
\begin{figure}[t]
 \center
\includegraphics[width=7.cm,height=9cm]{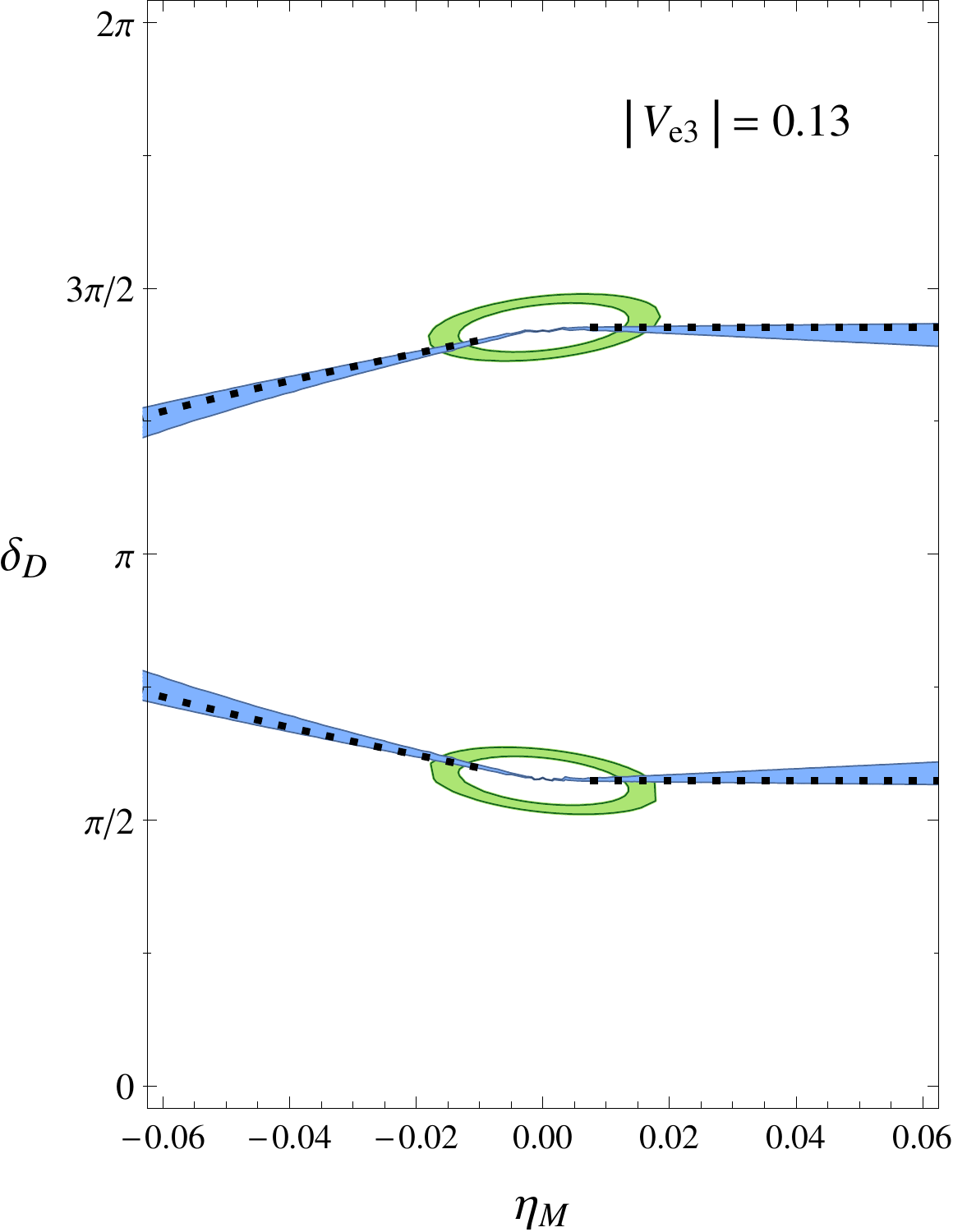}\hspace{1cm}
\includegraphics[width=7.cm,height=9cm]{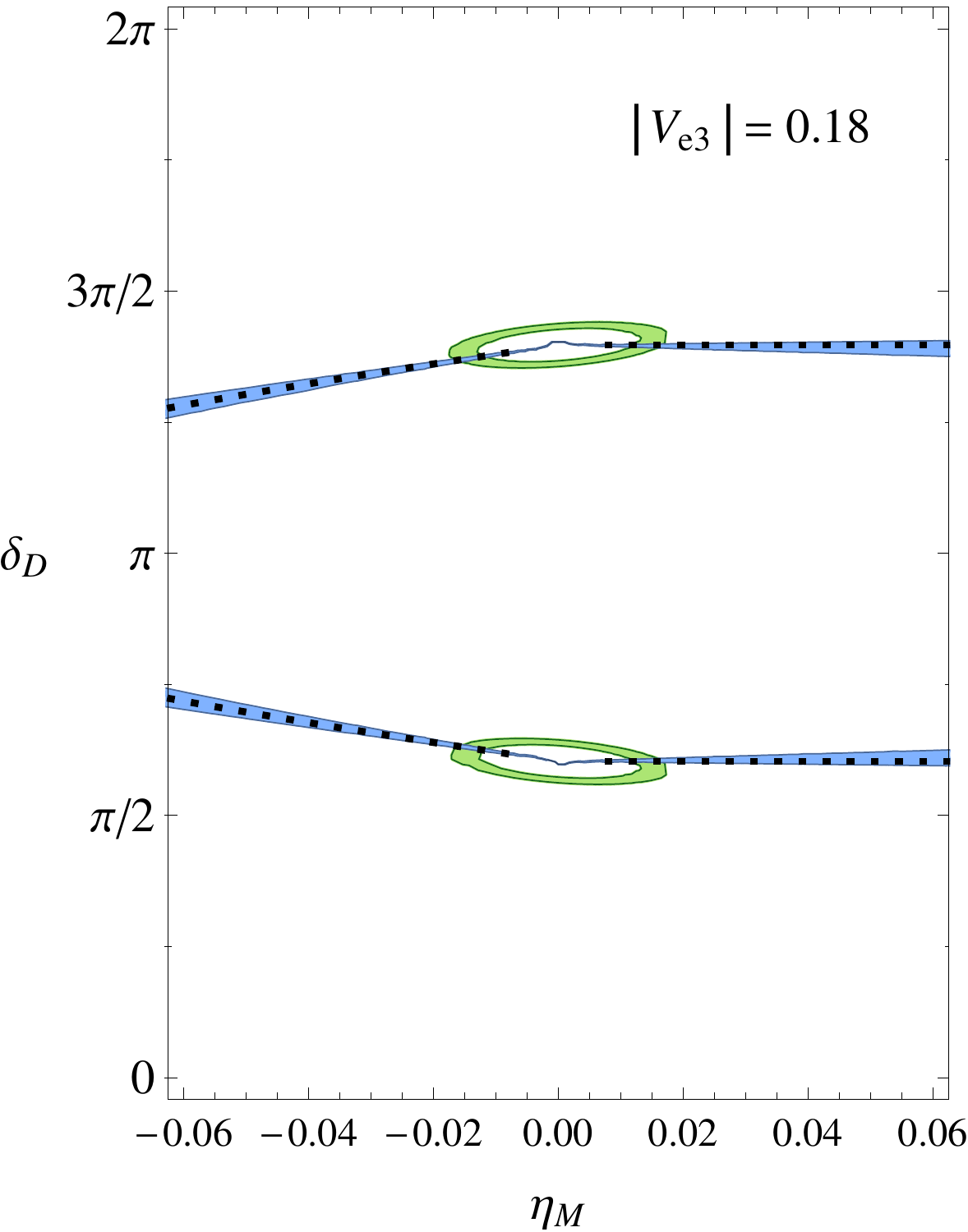}\hspace{1cm}
 \caption{Contours in the plane $(\delta_D,\eta_M)$ (where
   $\eta_M$ is such that $\displaystyle M_2=\frac{M_3}{2}(1+\eta_M)$) showing the regions where $0.509<|V_{e2}|<0.582$
   (blue bands) and where the neutrino mass ratio  $r=\frac{\Delta m_{21}^2}{\Delta m^2_{13}}$ is
   such that $0.0264<r< 0.036$ (green ellipses). In the left panel we fix $|V_{e3}| =0.13$ and
  in the right panel $|V_{e3}|=0.18$. In both panels, the dotted lines
  represent the approximation of Eqs.~(\ref{cos1}) and (\ref{cos2}).}
\label{contplot}
\vspace{.2cm}
\end{figure}
This fact pushes us to try and make further approximate analytical
predictions in order to obtain a simple expression for the viable values of $\delta_D$ in
this ansatz. Since we observe in Figure 2 that in the viable region of
parameter space $r \simeq 2\ |\eta_M|$, we will use this approximation
in Eq.~(\ref{ve2}) and enforce the tri-bimaximal value
$|V_{e2}^{tb}|^2=\frac{1}{3}$ as a first order approximation.
We
obtain the following constraints on the value of the CP violating phase $\delta_D$,
\bea
\cos{\delta_D}&\simeq& -\frac{5}{2\sqrt{2}}|V_{e3}| \hspace{3cm} (\eta_M>0) \label{cos1}\\
\cos{\delta_D}&\simeq& -\frac{5}{2\sqrt{2}}|V_{e3}| -
\frac{2}{3}\frac{\eta_M}{|V_{e3}|}\hspace{1.3cm} (\eta_M<0) \label{cos2}
\eea
These approximations appear in Figure 2 in the form of dotted
curves, and it is apparent that they fit the numerical results extremely
well. This tight prediction of the Dirac phase $\delta_D$ as a
function of $|V_{e3}|$ (along with the prediction of an inverted
spectrum) is a most important element of the ansatz as it can be
easily falsified as new neutrino data and global fits further tighten
the bounds on leptonic CP violation.

Finally, we compute the  rephasing invariant quantity defined as $J =
Im \{V_{e2} V_{\mu 3}V^*_{e3}V^*_{\mu 2}\}$, which is a  measure
${\cal CP}$ violation. In  our Ansatz it is given by

\begin{equation}
J \simeq \frac{1}{3\sqrt{2}}|V_{e3}|\sin(\delta_D)
\end{equation}
where we have used $2|\eta_M| \simeq r$ which is observed to fit nicely in the neighborhood of the tri-bimaximal texture.

\section{Leptogenesis and Neutrinoless Double Beta Decay}
Now, we will discuss leptogenesis in the present model.  For that we
will assume that in early universe,  the heavy Majorana neutrinos,
$N_i$, were produced  via scattering processes and  reached thermal
equilibrium at   temperature   higher than the see-saw scale. Since
the    mass term $N_i N_i$   violates the total lepton number by two
units,  the out of equilibrium decay of  the right handed (RH )
neutrinos\footnote{We will work in the basis where the mass matrix
  $M_{R}$ is a diagonal matrix.}  into the standard model leptons
can be a natural source of lepton asymmetry \cite{FY}.  The CP
asymmetry due to  the decay of $N_i$ into a  lepton with  flavor
$\alpha$ reads
\begin{eqnarray}\
  \epsilon_i^{\alpha}    = \frac{1}{8\pi v^2}\sum_{j\neq i}\frac{Im
    \left[(m^{+}_Dm_D)_{ij}\left(m^{+}_D\right)_{i\alpha}\left(m_D\right)_{\alpha
        j}\right]}{(m^{+}_Dm_D)_{ii}}F(M_i, M_j)
\end{eqnarray}
where   $F(M_i, M_j) $ is  the function   containing the one loop
vertex and self-energy corrections \cite{Loop}. For heavy neutrinos
far from almost degenerate its expression  is  given by
\begin{eqnarray}
F (M_i, M_j) =  \frac{M_j}{M_i} \left[\frac{M^2_i}{M^2_i - M^2_j} + 1 - \left(1 + \frac{M^2_j}{M^2_i}\right) \ln{\left(1 + \frac{M^2_i}{M^2_j} \right)}\right]
\end{eqnarray}

As the temperature of the universe cools down to about $100~GeV$, sphaleron
processes \cite{KRS}  convert the lepton-anti-lepton asymmetry  into a
baryon asymmetry \cite{BPY}. If one takes into  account the  flavor
effects,  and assume that the CP  asymmetry is dominated by $N_1$,
then  there are three regimes for the generation of the baryon
asymmetry \cite{Flavor-Lepto} (see also \cite{DNN}):
\begin{eqnarray}\label{etab}
|\eta_B| \simeq \left\{ \begin{array}{ll}
\!\!  1\times10^{-2} \sum_{\alpha =e,\mu, \tau}{\epsilon^{\alpha}_1} W\left(\tilde{m_1}\right) ;
 \hspace{6cm}  \ \ (M_1 \geq 10^{12} {\rm GeV})  \vphantom{\int^\int_\int}\!\!\!\!\!\!\!\!\!\!\!\\
\!\!3\times10^{-3}(\epsilon^{e}_1 +
  \epsilon^{\mu}_1)W(\frac{417}{589}(\tilde{m}^{e}_1  +
  \tilde{m}^{\mu}_1 )) +
  \epsilon^{\tau}_1W\left(\frac{390}{589}(\tilde{m}^{\tau}_1
  )\right);\   (10^9 {\rm GeV}\!\leq\! M_1\! \leq\! 10^{12} {\rm
    GeV} \vphantom{\int^\int_\int} )\!\!\!\!\!\!\!\\
\!\!3\times 10^{-3}\epsilon^{e}_1
  W\left(\frac{151}{179}(\tilde{m}^{e}_1 )\right) +
  \epsilon^{\mu}_1 W\left(\frac{344}{537}(\tilde{m}^{\mu}_1
  )\right)   +
  \epsilon^{\tau}_1W\left(\frac{344}{537}(\tilde{m}^{\tau}_1
  )\right);\ \ \ \ \  (M_1 \leq 10^{9} {\rm  GeV})\!\!\!\!\!\!\!\!\!\!\!&
\end{array} \right.
\end{eqnarray}
%
\begin{figure}[t]
 \center
\includegraphics[width=8cm,height=9cm]{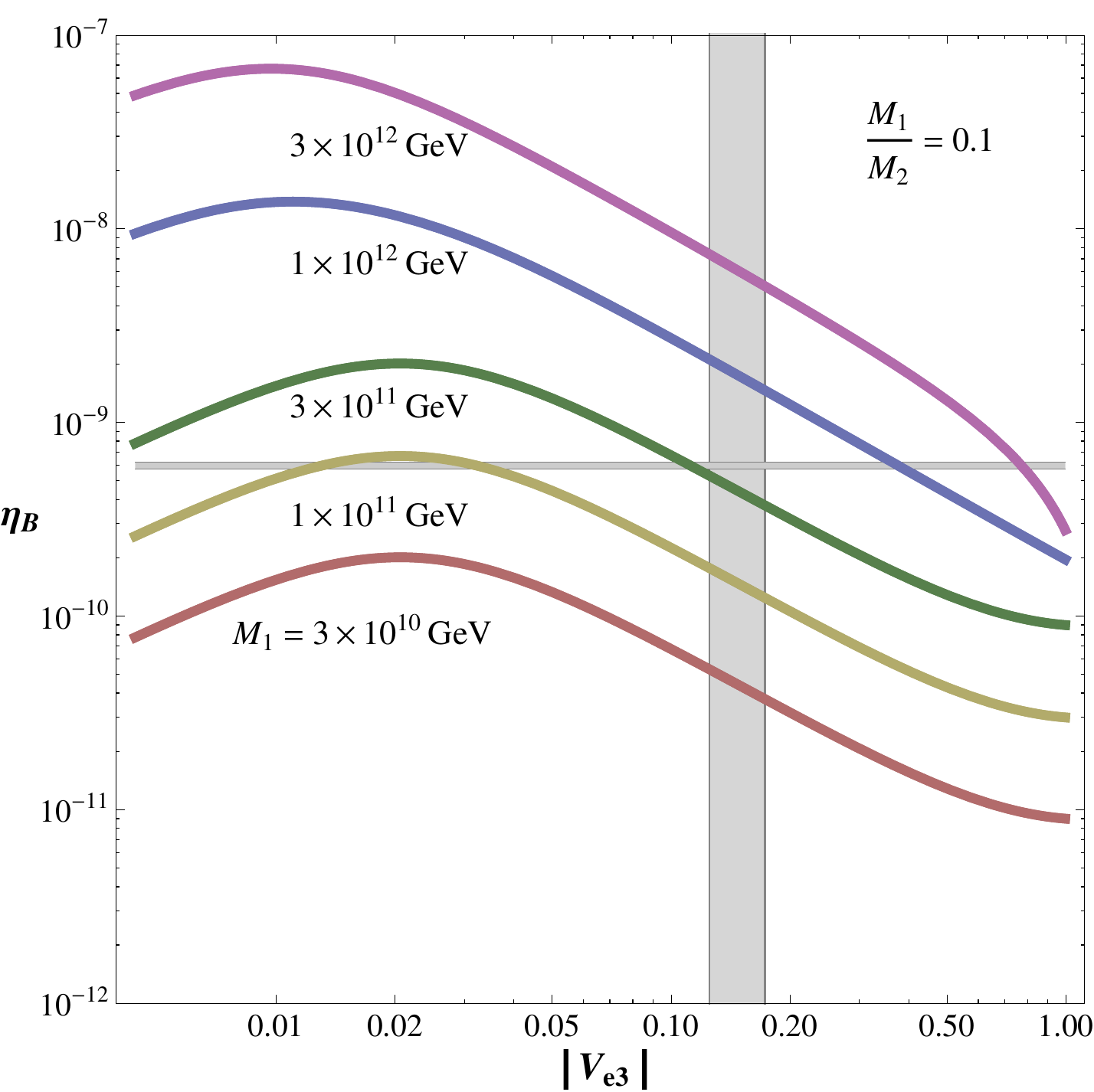}
\includegraphics[width=8cm,height=9cm]{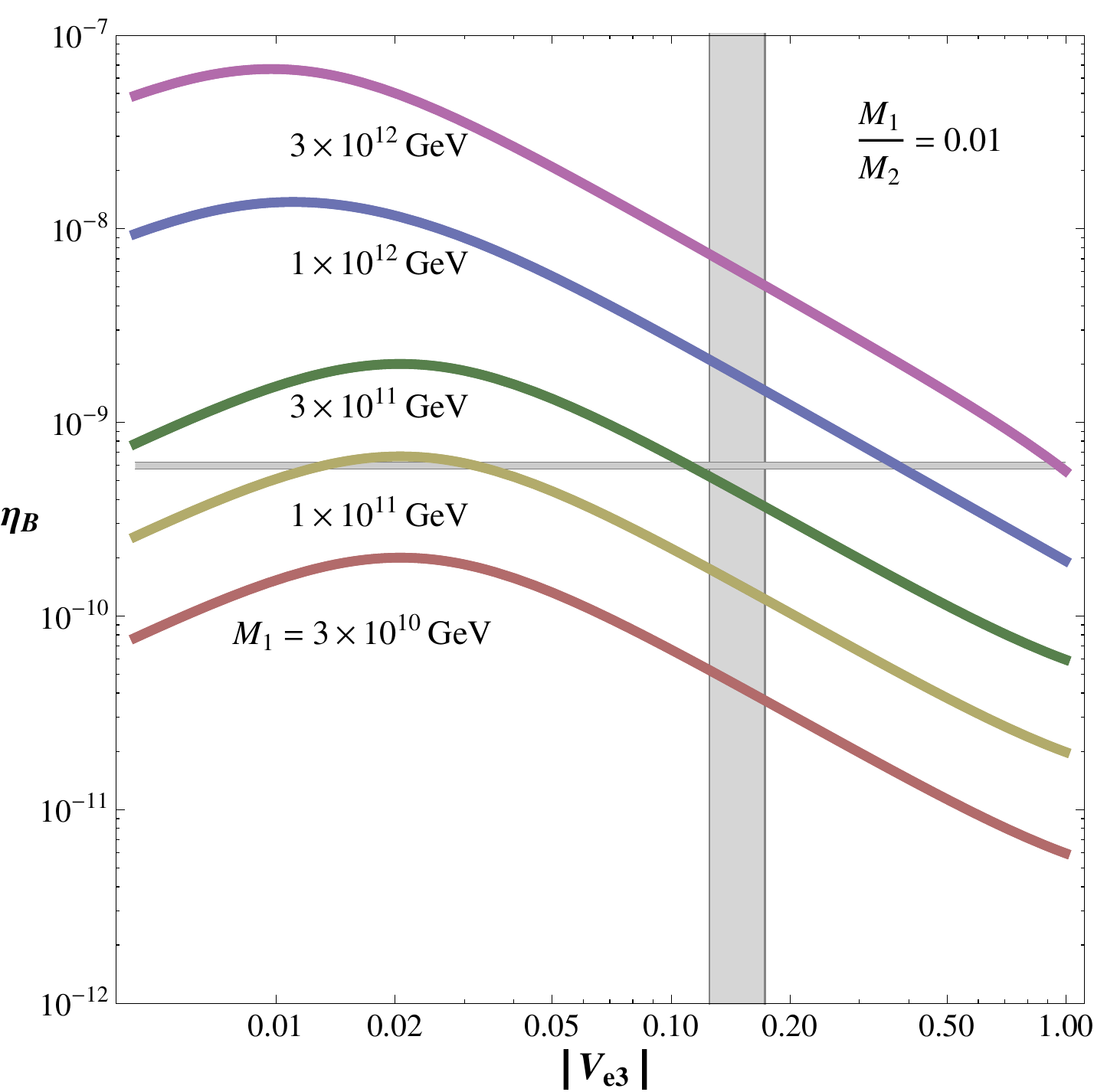}
 \caption{Baryon asymmetry produced in our specific scenario as
 a function of $|V_{e3}|$, in a hierarchical limit for
 the masses of the two lightest heavy Majorana
 masses, i.e. $M_1/M_2=0.1$ and $M_1/M_2=0.01$  .
The horizontal and vertical  bands represent the current experimental
bounds on $|\eta_B|$  and $|V_{e3}|$. Interestingly, we observe that
the higher the value of $|V_{e3}|$, the higher the required mass of
$M_1$ necessary to generate enough baryon asymmetry.
 }
\label{leptoplot}
\vspace{.2cm}
 \end{figure}
where
\begin{eqnarray}
\tilde {m_i}&=& \frac{\left(m^{+}_D m_D\right)_{ii}}{M_i} ; \\
\tilde{m}_i^{\alpha}&=&   \frac{\left(m^{+}_D\right)_{i\alpha} \left(m_D\right)_{\alpha i}}{M_1} ; \;\;\; \alpha =e,  \mu,  \tau\\
W(x) &\simeq& \left[\frac{8\times 10^{-3} eV}{x} + \left(\frac{x}{2\times 10^{-4} eV}\right)^{1.16}\right]^{-1};
\end{eqnarray}
Note that  in  the above  expressions of $\tilde{m}_i$ and
$\tilde{m}_i^{\alpha}$ there is no summation over repeated
indices. The quantity $W(x)$ accounts for the washing out of the total
lepton asymmetry due to $\Delta L = 1$ inverse decays.
If there is a strong hierarchy between the heavy neutrino masses, i.e.   $M_1
\ll M_2 \ll M_3 $, the asymmetry is dominated by the out of
equilibrium decay   of  the lightest one, $N_1$,  with $F (M_1,
M_{j\neq 1}) \simeq -\frac{3}{2}R_{1j}$. In this case, by using the
expressions of the mass matrix $M_D^{\dagger}M_D$:
\begin{equation}
M_D^{\dagger}M_D =|m_D|^2\left(\begin{array}{ccc}
3|z|^2 & z^* & z^* \\
z & \frac{3}{2} & 0 \\
z & 0 & 3 \end{array}\right)
\end{equation}
we find that the individual lepton flavor asymmetries are given by
\begin{eqnarray}\label{leptoapprox}
  \epsilon_1^{e}  &\simeq & \frac{M_1|\tilde{m}_\nu|(3+\eta_M)\sin(\delta_D)}{48\pi v^2}\\
 \epsilon_1^{\mu} &=& -   \epsilon_1^{\tau} \simeq  -\frac{M_1|\tilde{m}_\nu| \eta_M \sin(\delta_D)}{48\pi v^2}
 \end{eqnarray}
Thus, the high energy CP asymmetry is directly proportional to the  CP
violating phase of the effective low energy theory of the neutrino
sector.
Note that in the present model, $\delta_D \simeq \pi/2$,  which  allows
for the possibility  that  CP violation  could be  observed in
neutrino ( and   anti-neutrino) long baseline oscillation experiments
\cite{CPV1, CPV2, CPV3}.
\\

For the case where  two of the RH  neutrinos, say $N_1$ and $N_2$, are
almost degenerate, then  the  function $F (M_i, M_j) $ is dominated by
the contribution of  the one loop self energy diagram  and it is
given by \cite{Resonant} \begin{eqnarray}\label{Resonant}
F(M_i, M_j) = -\frac{\Delta M^2_{ij}M_iM_j}{\left(\Delta
  M^2_{ij}\right)^2 +  M^2_i\Gamma_i^2};~~~~~~~i, j = 1, 2 \end{eqnarray}
Here  $\Delta M^2_{ij} = \left(M^2_j - M^2_i\right)$ and $\Gamma_i =
\left(m^{+}_Dm_D\right)_{ii}/{8\pi v^2} M_i$ is the decay width of the
$i^{th}$ right-handed neutrino.  As a result,  the lepton  asymmetry
produced from  the decay of $N_1$ and $N_2$ can be considerably enhanced
when the mass splitting is of the order of the decay width of
$N_{1,2}$.    In the strong wash-out regime, the baryon asymmetry can
be estimated   using the  analytic expression\cite{BFJN,DP}\footnote{ In
  equation (61)  in  reference  \cite{ABD}, the expression of the baryon
  asymmetry for $M_1 \simeq M_2$ and  without considering the flavor
  effect is   approximated as
\begin{eqnarray}\nonumber
\eta_B \simeq -   10^{-2}  \sum_{\alpha =e, \mu, \tau}{\left(
  \epsilon^{\alpha}_1 + \epsilon^{\alpha}_2\right)
  \kappa_{\alpha}\left ( K^{\alpha}_1 +   K^{\alpha}_2\right)}
\end{eqnarray}
where $\kappa_{{\alpha}}$ is  the wash-out factor is  given by
\begin{eqnarray}\nonumber
\kappa_{\alpha} (x) \simeq \frac{2}{\left(2 +4x^{0.13}~ e^{-2.5/x}\right) x}
\end{eqnarray}
which is valid in the limit where $N_1$ and $N_2$ are   almost
degenerate \cite{BD}.  We have checked that the plots of the baryon  asymmetry obtained using this expression  agree well with the one presented in Fig.~4. }
\begin{eqnarray}
\eta_B \simeq - 2.4 \times 10^{-2} \sum_{\alpha =e, \mu, \tau} { \frac{ \sum^2_{i =1}{\epsilon^{\alpha}_i} }{ \sum^2_{i=1} K^{\alpha}_i \ln{(25 K^{\alpha}_i})}}
\end{eqnarray}
where
\begin{eqnarray}
K^{\alpha}_i =  \frac{\Gamma (N_i \rightarrow L_{\alpha} + H^{\dagger}) + \Gamma (N_i \rightarrow \bar{L}_{\alpha} + H)}{\zeta (3)H_{N_i}} \simeq \left(\frac{\tilde{m}^{\alpha}_i}{10^{-3}~eV}\right)
\end{eqnarray}
with $H_{N_i} \simeq 1.66 \sqrt{g_{*}} M_i^2/{M_{Pl}}$ is the Hubble
parameter at  temperature $T = M_i$ , where $M_{Pl} =  1.2 \times
10^{19}~GeV$ is the planck mass, and $g_{*} = 106.75$ is the total
number of degrees of freedom.  Here the asymmetries
$\epsilon^{\alpha}_i$ are calculated using the expression of the
function $F (M_i, M_j)$ given in Eq (\ref{Resonant}).
\\

\begin{figure}[t]
 \center
\includegraphics[width=8cm,height=9cm]{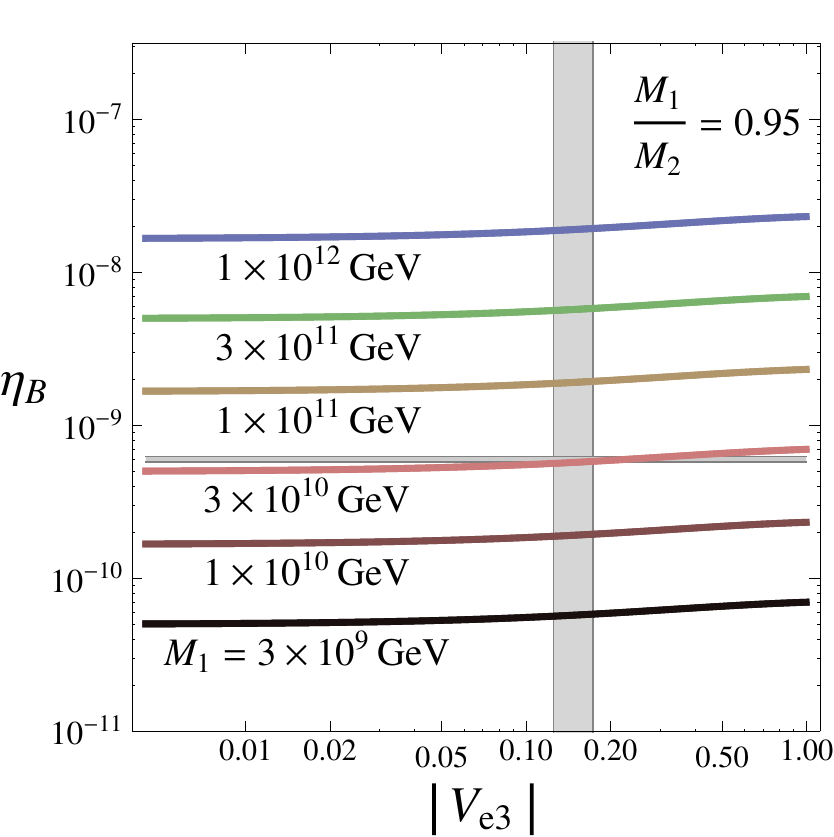}
\includegraphics[width=8cm,height=9cm]{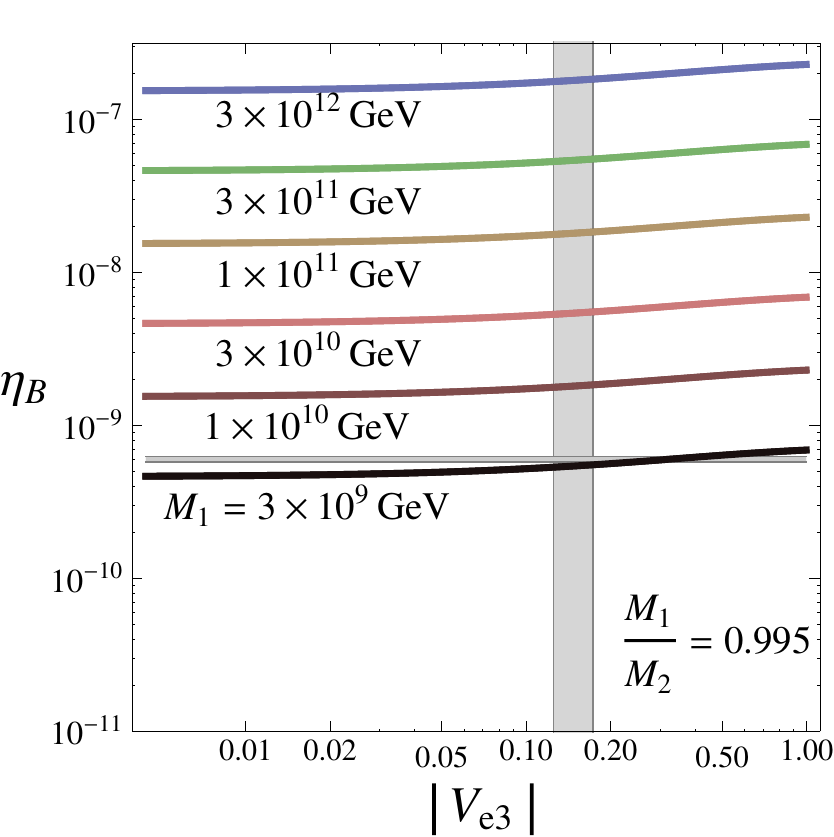}
 \caption{Baryon asymmetry produced in our specific scenario as
 a function of $|V_{e3}|$, in a limit in which the two lightest heavy
 Majorana masses are nearly degenerate, i.e. $M_1/M_2=0.95$ and
 $M_1/M_2=0.995$, thus producing a resonant enhancement of the asymmetry. The
 horizontal and vertical  bands represent the current experimental
 bounds on $|\eta_B|$ and $|V_{e3}|$. Note that the dependence on $|V_{e3}|$
 is much milder than in the non-degenerate case.
 }
\label{leptoplot}
\vspace{.2cm}
 \end{figure}

We show in Fig.~3 the dependence of the baryon asymmetry on the
reactor mixing  parameter $|V_{e3}|$ for  different values of $M_1$,
ranging  from  $3\times 10^{10}~GeV$ to $3 \times 10^{12}~GeV$ with
$R_{12} = 0.1$ and $R_{12} = 0.01$ (hierarchical mass
limit). We see  that successful leptogenesis requires
that   $M _1 \simeq 3 \times 10^{11}~GeV$, and also that there is an
interesting dependence on $|V_{e3}|$, due to flavor effects, such that smaller values
correspond to higher asymmetry. Irrespective of the experimentally allowed  values of
$|V_{e3}|$, we find that for  $M_1 \leq 10^{11}~GeV$,  the  value of
$\eta_B$ is too small to account for the observed matter-anti  matter
asymmetry of the universe, due to the strong wash-out effect.
In Fig. 4, we make a similar plot  for the case of  almost degenerate right handed neutrino spectrum, where we  consider $R_{12} = 0.95$ (left panel) and $R_{12} = 0.995$ (right panel). It shows that it is possible to to generate a baryon asymmetry in agreement with the observation for $M_1$ smaller than $10^{11}~GeV$,  thanks to the
     resonant effect when the masses of $N_1$ and $N_2$ are sufficiently close. In that limit, the flavor effects are now
     different and indeed we observe that the dependence on $|V_{e3}|$ is
     much milder obtaining basically flat curves, whose heights are
     increased for values of $R_{12}$ closer to $1$.  For instance,  when $R_{12}=
     0.95$,  a RH neutrino with  mass  $M_1 \sim 3\times 10^{10}~GeV$ can  produce the correct baryon asymmetry. If  the degeneracy between $M_1$
     and $M_2$ is made stronger, as  for  our choice  of    $R_{12}=0.995$, the
     mass for $M_1$ is lowered  by an order of magnitude to $M_1 \sim 3\times 10^{9}~GeV$.
\\

Now, we compute the contribution to the effective mass
$m_{\beta\beta}$ which parameterizes the neutrinoless double beta
Decay. Note that $m_{\beta\beta}=|S_{11}|$, with $S_{11}$ is given by Eq.~(\ref{Snu}).

\begin{eqnarray}
m_{\beta\beta}^2 \simeq |\Delta m^2_{13}|
\left[1+\frac{|V_{e3}|\cos(\delta_D)}{\sqrt{2}}+\frac{5|V_{e3}|^2}{4}+\frac{r}{3}\right],
\end{eqnarray}

where we have used the following expansion for $|\tilde{m}_\nu|^2$
(making use of the approximation $\displaystyle \eta_M \simeq \frac{r}{2}$),

\begin{eqnarray}
|\tilde{m}_\nu|^2 \simeq |\Delta
m^2_{13}|\left[1-\frac{|V_{e3}|\cos(\delta_D)}{\sqrt{2}}+\frac{3|V_{e3}|^2}{4}\right]
\end{eqnarray}
Since in this model, the Dirac CP phase is approximately $\pi/2$, we can write
\begin{eqnarray}
m_{\beta\beta} \simeq \sqrt{|\Delta m^2_{13}|}\left(1+ \frac{5|V_{e3}|^2}{8}+\frac{r}{6}\right),
\end{eqnarray}
Thus, for the mass texture (\ref{MinText}), neutrinoless double beta
mass parameter is predicted  to be $m_{\beta\beta} \simeq 5 \times
10^{-2}~eV$, which is smaller than the current bound by about an
order of magnitude.  However,   experiments such as GERDA, CURO, and
MAJORANA  with 1 ton.yr exposure will have sensitivity of about
$0.03~e V$\cite{DBD}, and hence  it will be possible to test the above
prediction.

\section{Conclusion}

In this paper we investigated some of the implications of deviating
from exact $\mu-\tau$ symmetry assuming that neutrino masses are
generated via the see-saw mechanism. A simple parametrization of the
Dirac neutrino mass matrix, $M_D$, with just $3$ parameters, was presented and
studied. The scenario is consistent with all neutrino oscillations data and has interesting
predictions for some of the observable parameters.
We were able to find transparent relations
among the different observables of the setup, and in particular the
value of the Dirac CP phase happens to be highly constrained as a function of the
mixing angle $V_{e3}$. The dependence of the other mixing angles of the $V_{PMNS}$
mixing matrix in terms of $V_{e3}$ was also obtained. The neutrino
masses are also linked directly to the see-saw structure in a very
simple way as well as the lepton asymmetry generated out of the decay of the
lightest right handed neutrino. We find that lepton asymmetry is directly proportional to the mixing
angle $|V_{e3}|$, which thus has to be non vanishing to be in agreement
with the observed baryon asymmetry of the universe. The Dirac phase
happens to be also the relevant phase for leptogenesis, linking low
scale CP violation to high scale CP violation in a transparent
way. Moreover the predicted value for the Dirac phase (close to $\pi/2$)
gives an almost maximal contribution to leptogenesis.

We expect that all the different types of ansatzes that can be considered
in our framework of partial $ \mu-\tau$ will have similar simple
predictions and structures as the one studied here.
A thorough investigation is underway and will be the subject of
future publication.

\section{Acknowledgements}

One of us (C.H.) would like to thank Zhi-Zhong Xing for useful
discussions and acknowledge the support and hospitality of the High Energy
Institute in Beijing. C.H. also wishes to thank Michel Lamothe for discussions..

\end{document}